\documentstyle[11pt,newpasp,twoside,epsf]{article}
\def\edcomment#1{\iffalse\marginpar{\raggedright\sl#1\/}\else\relax\fi}
\reversemarginpar

\begin{document}
\title{What are MACHOs?  Interpreting LMC microlensing}
 \author{David S. Graff}
\affil{The Ohio State University\\
 Departments of Physics and Astronomy\\
 Columbus, OH 43210 USA}

\begin{abstract} 

I discuss two hypotheses that might explain LMC microlensing: the Halo stellar remnant lensing hypothesis and the unvirialized LMC lensing hypothesis.  I show that white dwarfs cannot contribute substantially to the cosmic baryon budget; they are strongly constrained by chemical evolution and background light measurements.  Although there have been some claims of direct optical detections of white dwarfs in the Halo, I show how the full sample of direct optical searches for halo lenses do not support the Halo lens hypothesis.

N-body simulations suggest that the LMC may be naturally excited out of virial equilibrium by tidal forcing from the Milky Way. New measurements of LMC kinematics not only do not rule out the unvirialized LMC lensing hypothesis, but even moderately favor it (at 95\% confidence).

\end{abstract}

\section{Introduction}

Microlensing today has a wide variety of scientific applications ranging from stellar atmospheres (Sackett, this volume) to Galactic structure (Binney, this volume) to searching for planets (Gaudi, this volume), but the initial prime mover of microlensing, which seduced so many particle astrophysicists and cosmologists such as myself into dirtying our hands with grungy astronomy, has been the search for dark matter in the Milky Way halo.  Specifically, we wish to determine what fraction of the halo is composed of lumps of dark matter that could lens background stars in other galaxies, such as the Magellanic Clouds.  These lumps of matter are known as MACHOs (MAssive Halo Compact Objects) by which I refer to any halo objects massive enough to cause detectable gravitational microlensing.  The MACHOs stand in contrast to WIMPs (Weakly Interactive Massive Particles), putative particle-dark matter candidates that are not massive enough to microlens.  MACHOs are also distinct from Halo stars, sometimes known as the stellar Halo.  Halo stars are massive enough to lens, but are not dark.  They make up less than 1\% of the mass of the Halo (Graff \& Freese 1996) and thus cannot be responsible for the microlensing events seen towards the LMC.

There are several candidate objects that in principle could be MACHOs; they are reviewed in Carr (this volume).

As dark matter experiments, microlensing searches have been phenomenally successful, among the most successful of all dark matter searches.  The MACHO and EROS experiments have published a combined limit in which they have ruled out enormous regions in parameter space (Afonso et al. 2000), limits that have been tightened by results presented at this meeting (Lasserre, this volume; Cook, this volume). These results exclude objects with mass in the range $10^{-7} - 1 M_\odot$ from making up all the dark matter in the halo.

However, microlensing results still present a mysterious signal; the MACHO experiment has reported an excess of microlensing events towards the Magellanic Clouds with time scales of tens of days.  These events are difficult to explain.  If one interprets them as being due to a halo of lenses, then their total mass is $9^{+4}_{-3} \times 10^{10} M_\odot$, which, depending on the halo model chosen, makes up some $\sim 20\%$ of the mass of the halo. They have long time scales which, if interpreted as being due to MACHOs in the Halo, suggest that the lenses have mass in the range $0.1 - 1 M_\odot$ (Cook, this volume).  The only known dark astrophysical objects that have masses in this range are white dwarfs.  Taken at face value, this result could be interpreted to mean that white dwarfs are common in the universe, perhaps being a significant fraction of all baryons.

The white dwarf hypothesis gained strength when Ibata et al. (1999) claimed to have detected them in the Hubble Deep Field and Ibata et al. (2000) claimed further detection of one additional halo white dwarf in a photographic proper motion survey.

In this paper, I will show that white dwarfs make poor dark matter candidates, and contribute negligibly to the cosmological baryon budget.  Thus, microlensing experiments have not identified a significant baryonic dark matter candidate.  I will show that the detections of Ibata et al. do not require a large population of halo white dwarfs.  

I have not been able to completely rule out white dwarfs as being responsible for microlensing; in fact, Gates (this volume) has suggested that the MACHO experiment may have detected a new component of the Galaxy composed of white dwarfs.  Still, I will establish that white dwarfs are strongly constrained, exotic lensing candidates requiring several conditions.

Before one accepts any potentially controversial interpretation of an experimental result, one must be sure that the result is not due to an unaccounted for background signal in the experiment.  I will discuss a possible background for the microlensing experiments, lensing by ordinary stars near the LMC.  I will show that this background has not been ruled out observationally, is suggested by some theoretical calculations, and has even been detected observationally, though the statistical significance of these detections is not compelling.

\section{Can MACHOs be White Dwarfs?}

\subsection{Cosmology and MACHOs}

We begin this section with a discussion of the total cosmological density of MACHOs in units of the critical density, $\Omega_{\rm macho}$.  Most of this work is taken from Fields, Freese, \& Graff (1998).  Since Microlensing has as its ancestry, the search for dark matter, the cosmological density of MACHOs is of critical importance, if anything, more important than the mass density of MACHOs around the Milky Way.

Fields et al. (1998) placed upper and lower limits on the Cosmological densities of MACHOs, which I here update to include the new MACHO results presented in Cook (this volume).  If we assume that MACHOs trace dark matter, then $\Omega_{\rm macho}=\Omega_m f_{\rm halo} \sim 0.06$.  Here, $\Omega_m$ is the density of cosmological density of collisionless matter and $f_{\rm halo}$ is the fraction of the mass of the halo within the LMC radius of 50 kpc composed of MACHOs.  Making the minimal assumption that only Milky Way type spiral galaxy halos contain MACHOs, the lower bound on the cosmic density of MACHOs is $\Omega_{macho}> 0.001 - 0.01$

The above discussion is independent of what the MACHOs actually are. What if the MACHOs were stellar remnants, white dwarfs or neutron stars?  Stellar remnants are in some respects attractive MACHO candidates; they have the right mass, and are dark.  However, as I will show, stellar remnants make poor dark matter candidates.

The constraints against stellar remnants are ultimately due to limits on the quantity of nuclear fusion.  Large numbers of white dwarfs mean that lots of nuclear fusion had to take place.  While a traditional stellar population will convert perhaps $\sim 5\%$ of its mass from hydrogen to helium, most of the hydrogen being locked up in dwarf stars, a stellar remnant will contain no hydrogen.  Pound for pound, a white dwarf MACHO population implies thus that $\sim 20$ times more fusion will have taken place than a similar mass of stars.

Since a large white dwarf MACHO population implies that there has been copious fusion, we can place limits on this population by placing limits on the two byproducts of fusion, energy and heavy elements. A cosmologically significant population of white dwarfs severely modifies the chemical enrichment of its surrounding.  Under a typical scenario, a 3 $M_\odot$ progenitor star will eventually die into a 0.6 $M_\odot$ white dwarf, but the remaining 80\% of its mass is released as chemically enriched gas.  The chemical evolution implied by this gas is analyzed in Fields, Freese \& Graff (2000, FFG00).

The strongest limits come from abundances of carbon and nitrogen.  Standard chemical yields of low metallicity stars suggest that the ejected gas of white dwarfs is enriched with a solar abundance of either carbon or nitrogen (depending on the extent of hot bottom burning).  Since Galactic halo stars have a carbon abundance of $\sim 10^{-2}$ solar, only $\sim 10^{-2}$ of halo material can have processed through stars at the time of the formation of the halo. Similarly, the mean carbon enrichment of the universe at $z\approx 3$ is $\sim 10^{-2}$ solar as measured in Lyman $\alpha$ absorption systems.  Thus we see that only $\sim 10^{-2}$ of all baryons can have passed through stellar progenitors of white dwarfs by $z\approx 3$. (See FFG00 for a more detailed calculation).

Chabrier (1999) has suggested that zero metallicity white dwarfs may not emit carbon, finessing these limits.  In that case, more robust, but less restrictive chemical evolution limits can be placed with Helium.  FFG00 showed that the helium evolution due to a cosmological remnant population limits Pop. III remnants to $\Omega_{WD}<0.002h^{-1}$, still an insignificant component of baryons.  Thus, remnants cannot contribute significantly to baryonic dark matter.

A parallel limit can be placed using the light emitted by the stellar progenitors.  This limit is slightly weaker than the chemical evolution limits, but far more robust.  Zero metallicity stars may not emit Carbon, but they certainly emit light!  Using limits on the Infrared background, Graff et al. (1999) showed that $\Omega_{wd}< 0.004 h$.  Again, we conclude that stellar remnants cannot contribute to baryonic dark matter.

The original scientific goal of microlensing was to identify baryonic dark matter.  As we have seen, we can place robust limits on the cosmological density of stellar remnants, the only astrophysical objects that could be responsible for the signal seen by the MACHO experiment.  Thus, no stellar object can contribute significantly to baryonic dark matter.

\subsection{White Dwarfs in the Milky Way Halo}

Even though white dwarfs cannot be significant components of cosmological dark matter, might they not still be present in the Milky Way halo?  Could not the Milky Way halo be an exceptional place in which much of the mass was composed of stellar remnants?  Gates (this volume) suggested such a model.  In her system, the lensing detected by the MACHO experiment is due to a new stellar remnant component of the Milky Way, not associated with the dark matter halo.

It is difficult to constrain the possibility that the Milky Way halo alone could contain large numbers of white dwarfs.  Obviously, limits on background light do not apply to our own halo, but only to distant galaxies.  Gibson \& Mould (1997) noted that such a halo of white dwarfs was inconsistent with measured low carbon abundances of halo stars.  However, FFG00 noted that a galactic wind driven by SN Ia from these white dwarfs could blow the carbon-enriched gas out of the halo.

A possible way to confirm the existence of a spheroid population of white dwarfs is to try to directly detect them.  If such a spheroid exists, then its white dwarfs are the most common type of star in the Milky Way.  Their detection is not trivial however, because they can be quite dim, $M_V \sim 18$, and it is difficult to separate them from the overwhelming background of brighter Milky Way stars and distant galaxies.

Before I begin discussion of direct detection of white dwarfs, I should mention that there has been a recent revolution in the theoretical study of white dwarf luminosities, colors, and cooling curves.  Before 1997, no one had calculated what white dwarf atmospheres were like cooler than 4000 K, roughly the temperature of the then coolest observed white dwarf.  However, a Macho population of white dwarfs would be older and thus likely cooler than disk white dwarfs.  The new theory of white dwarfs is discussed in Hansen (this volume).

Two results from this analysis pertinent to optical searches for white dwarfs are as follows.  1) Old white dwarfs with helium dominated atmospheres would have cooled to invisibility, and could never be directly detected, and 2) cool hydrogen atmosphere white dwarfs emit most of their light in the V and R bands, and have spectral energy distributions very far from black body.

Point 1) means that no direct search for white dwarfs could ever rule out the existence of a Macho population of white dwarfs: the Machos could be all helium atmosphere white dwarfs.  However, Point 2) above makes optical searches for hydrogen atmosphere white dwarfs relatively more powerful and model independent, since the white dwarfs are emitting their light in optical frequencies.

There have been two broad strategies to search for search for a halo population of white dwarfs.  In one, deep images are taken at high galactic latitude, thus eliminating the background of disk stars. Here, the strongest background is distant galaxies.  The high spatial resolution of the HST can be used to separate galaxies from stars. Flynn, Gould \& Bahcall (1996) used this method with the Hubble Deep Field, and did not find evidence of white dwarfs.

The other method is to look for high proper motion objects.  Halo white dwarfs should have high proper motions due to their high velocity, and low intrinsic magnitude.  Proper motion searches have ranged from shallow photographic searches over wide solid angles (Luyten 1979, Ibata et al. 2000) to deep narrow searches of the Hubble Deep Field (Ibata et al. 1999).

Interpreting the proper motion surveys is controversial, since they are not consistent.  Flynn et al. (2000) compared the various proper motion surveys.  We found that the LHS survey (Luyten 1979) was by far the most powerful, tens of times more powerful than the other published surveys.  Yet, the LHS survey does not find evidence of a MACHO population of white dwarfs while two other surveys, Ibata et al. (1999) and Ibata et al. (2000) find a handful of objects suggesting that the halo might be full of white dwarfs.

There are two different possible interpretations of these conflicting results: either the halo is full of white dwarfs and the LHS survey does not see them, or the halo does not contain many white dwarfs, and the handful of objects seen by the two Ibata et al. surveys are a mix of Poisson fluctuations and background objects.  I consider these two possibilities below.

The LHS survey was done some thirty years ago, and its high proper motion stars were often detected by hand.  Thus, there is no modern artificial star--Monte Carlo estimate of its efficiency.  However, in Flynn et al. (2000), we estimated the efficiency of the LHS survey by the numbers of bright and dim stars, and estimated that it was at least 60\% efficient down to mag. $R_L=18.5$.  This estimate can only be based on dim, low proper motion start since there are very few dim high proper motion stars in the sample.  Possibly, the survey misses the dim high proper motion stars, although there are some reasons to think otherwise, which are discussed in Flynn et al (2000).

The Ibata et al. (1999) survey looked for proper motion of Hubble Deep Field objects; with the philosophy that anything that moved must be a star, and not a background galaxy.  However, in Graff \& Conti (2000), we show that Ibata et al. under estimated their proper motion uncertainties.  Thus, the objects they found are most likely distant galaxies.

The Ibata et al. (2000) survey used scanned Schmidt plates to search for high proper motion stars.  One advantage of this technique is that the objects found are close enough to be observed spectroscopically.
They found two dim, low temperature, spectroscopically confirmed white dwarfs, one of which was also in the LHS survey.

In their paper, Ibata et al. (2000) suggest that these objects imply that white dwarfs make up 10\% of the local mass density of the
Galactic halo.  This number is highly uncertain.  Just from Poisson statistics of 2 objects, the 90\% confidence lower limit is 4 times lower than the detected value, 2\% of the local mass density of the halo.  Ibata et al. also made poorly constrained assumptions about the mass of their objects and the absolute magnitude of the new white dwarf in their sample, the one that is not in the LHS survey.  Thus the real lower limit based on their observations is below 1\% of the local mass density of the halo.

This limit is low, comparable to the local density of halo stars. Thus, these two white dwarfs could simply be representatives of the local Pop II halo white dwarf population, which likely make up a bit less than 1\% of the local mass density of the halo.  It is still premature to say whether this survey has found evidence of a Macho population of white dwarfs.

\subsection{Conclusion}

 White dwarfs cannot make a significant component of Baryonic Dark Matter.  They would cause too much enrichment of carbon, nitrogen, and helium.  They would also create an infrared background above current limits.

 Even though white dwarfs cannot greatly contribute to the cosmic baryon census, they could be over-represented in the Milky Way halo. Direct searches can never rule out this possibility since a high fraction of white dwarfs could have helium atmospheres and be too dim to see.  There have been some claimed detections of white dwarfs, but, in my opinion, these detections are not yet compelling.

\section{Other lensing candidates, especially LMC self lensing}

 Various authors have proposed several exotic, non-baryonic lensing candidates that are reviewed by Carr (this volume).  However, before we are driven to these candidates, we must first verify that there are no other viable baryonic lensing candidates.

One obvious baryonic candidate is lensing by ordinary stars.  Ordinary stars have the right mass, typically $0.1-1 M_\odot$, but they are not usually thought of as making good dark matter candidates, since they are not Dark!  Ordinary stars are known to make up a negligible fraction of the local halo mass density (Graff \& Freese 1996).  However, various schemes have been proposed in which stars do cause a large optical depth towards the LMC.

Wu (1994) and Sahu (1994) proposed that the LMC could generate self lensing, lensing of LMC stars by other LMC stars.  However, Gould (1995) examined the self lensing optical depth of a virialized disk galaxy and showed it was proportional to the velocity dispersion, $\tau=2\sigma^2/c^2 \, sec^2 i$, where $\sigma$ is the line of sight velocity dispersion and $i$ is the inclination angle.  The velocity dispersion of the LMC is measured to be low in a variety of stellar populations representing a wide variety of ages and metallicities (reviewed by Gyuk, Dalal \& Greist 1999). Thus, the bulk of stars in the LMC are in a thin, face on disk with a low optical depth, lower than that observed by the MACHO group.

Zhao (1998) suggested that the Milky Way halo could contain small recently accreted objects such as the Sagittarius Dwarf, and tidal tails ripped off these objects and the Magellanic Clouds.  He proposed that if one of these lumps of matter were interposed along the line of sight to the LMC, it could perhaps cause sufficient microlensing to account for the MACHO observations.
 
If such systems of objects were distributed randomly in the Halo, as we would expect if they arose from Sagittarius-Dwarf-like systems, then the probability that an individual one would by chance be aligned with the LMC is quite low, $(\sim 10^{-4})$ (Gould 1999).  Yet they cannot be numerous, since if such objects were composed of ordinary luminous stars, in order to have a density sufficient to cause the measured microlensing, they would have a surface brightness high enough to be visible to the naked eye.  The only way they could have escaped detection is if they lie exactly in front of the LMC, or, like the Sagittarius dwarf, lay behind the galactic center.  Thus, the only likely way that there could be an object along the line of sight to the LMC is if it was somehow associated with the LMC, perhaps a tidal tail lifted off the LMC by interactions with the Milky Way or SMC, or perhaps a part of the LMC itself, a thick ``shroud'' (Evans \& Kerins 2000).

Several papers have been written about whether or not there is an object along the line of sight to the LMC, too many to review here.  They are discussed in detail in Zaritsky et al. (1999), whose basic conclusion is that such a population {\it cannot be ruled out} if it is close enough to the LMC in both positions and velocities, within approximately 10 kpc and 30 km s$^{-1}$.  If the object is in this range, it is swamped by stars within the LMC proper, and is difficult to detect.

In Graff et al. 2000, we searched for a kinematically separate population from the LMC using Carbon Stars.  The idea behind this search was that any separate population should have a velocity that is different from that of the LMC.  We found, at 95\% statistical confidence, evidence of a population displaced 30 km s$^{-1}$ from the LMC.  If this population exists, it would likely be at a distance from the LMC sufficient to cause a significant microlensing signal.

Numerical simulations of the LMC-Milky Way system support the idea that a chunk of the LMC was ripped off by the Milky Way.  If such a chunk lay along the line of sight, it would cause enough Microlensing to explain the MACHO group results.  In unpublished work, Mark Galpin and I have analyzed the N-body simulation of Weinberg (2000).  We found that the tidal features of this simulation are large enough to cause a microlensing optical depth of $1\time 10^{-7}$, consistent with the MACHO group results, along some lines of sight.

\subsection{Conclusion}

A yet undetected population of ordinary stars could cause the measured microlensing population.  This population must be close to the LMC in both distance and radial velocity, otherwise it would have already been detected.  N-body simulations suggest that such a population could have been ripped off the LMC by tidal interactions with the Milky Way.  Graff et al. (2000) have found, at only the 94\% confidence level, evidence of a non-LMC disk population, the Kinematically Distinct Population, or KDP.  If this population is confirmed, it could explain the Microlensing detected by the MACHO collaboration.

\section{Editorial Musings}

I have discussed two leading explanations for LMC microlensing events, that the halo is full of white dwarfs, or that the LMC is sufficiently far from virial equilibrium to generate a large optical depth with a small velocity dispersion.  To some extent, both explanations have similarly weak observational support; neither be ruled out, and each has statistically weak observational evidence.  However, the white dwarf hypothesis requires a chain of unlikely events: massive amounts of baryons must have undergone star formation with a radically different, extremely narrow mass function, and these stars must not have emitted any Carbon.  Further, this star formation mechanism can only have occurred in the inner halos of Milky Way type galaxies, from which a galactic wind must remove ejected gas.  Occam's razor suggests that before we believe such a long chain of new cosmology, we should adopt the simpler notion that microlensing experiments have simply found a background of lensing by ordinary stars.

\end{document}